\documentclass[prd,aps,nofootinbib,twocolumn,superscriptaddress,preprintnumbers,balancelastpage,longbibliography]{revtex4-1}
\usepackage{ae,aecompl}
\usepackage{graphicx}	
\usepackage{amsmath}	
\usepackage{graphicx}
\usepackage{dcolumn}
\usepackage{bm}
\usepackage{graphics}
\usepackage{afterpage}
\usepackage{float}
\usepackage{subfigure}
\usepackage{rotating}
\usepackage{multirow}
\usepackage{tabularx}
\usepackage{booktabs}
\usepackage{multirow}
\usepackage{fancyheadings}
\usepackage{fancyhdr}

\usepackage{theorem}
\usepackage{moreverb}
\usepackage{euscript}
\usepackage{psfrag}
\usepackage{slashed}
\usepackage{mathtools}
\usepackage{makecell}
\usepackage[flushleft]{threeparttable}
\usepackage{adjustbox}
 
\usepackage[utf8]{inputenc}
\usepackage[english]{babel}

\usepackage{dcolumn}
\usepackage{bm}
\usepackage[mathlines]{lineno}

\usepackage[colorlinks=true
,urlcolor=blue
,anchorcolor=blue
,citecolor=blue
,filecolor=blue
,linkcolor=black
,menucolor=blue
,linktocpage=true
,pdfproducer=medialab
,pdfa=true
]{hyperref}
\usepackage{listings}
\usepackage{color,xcolor}

\newcolumntype{C}[1]{>{\centering\let\newline\\\arraybackslash\hspace{0pt}}m{#1}}

\lstdefinestyle{python}{
  belowcaptionskip=1\baselineskip,
  breaklines=true,
  frame=L,
  xleftmargin=\parindent,
  language=Python,
  showstringspaces=false,
  basicstyle=\small\ttfamily,
  morekeywords={models, lambda, forms,True,False,None},
  keywordstyle=\bfseries\color{deepgreen!40!black},
  commentstyle=\itshape\color{gray},
  identifierstyle=\color{black},
  stringstyle=\color{deepred},
  rulecolor=\color{gray},
}

\begin{document}

\title{Inflation Wars: A New Hope}

\author{Ryan E.~Keeley}
\affiliation{Korea Astronomy and Space Science Institute, Daejeon 34055, Korea}

\author{Arman Shafieloo}
\affiliation{Korea Astronomy and Space Science Institute, Daejeon 34055, Korea}
\affiliation{University of Science and Technology, Daejeon 34113, Korea}
\author{Dhiraj Kumar Hazra}
\affiliation{The Institute of Mathematical Sciences, HBNI, CIT Campus, Chennai 600113, India}
\affiliation{Osservatorio di Astrofisica e Scienza dello Spazio di Bologna/Istituto Nazionale di Astrofisica, via
Gobetti 101, I-40129 Bologna, Italy}
\affiliation{Istituto Nazionale Di Fisica Nucleare, Sezione di Bologna,Viale Berti Pichat, 6/2, I-40127 Bologna, Italy}
\author{Tarun Souradeep}
\affiliation{Indian Institute of Science Education and Research, Pune, India}
\affiliation{Inter-University Centre for Astronomy and Astrophysics, Pune 411007, India}

\date{\today}

\newcommand{\aap}{Astronomy and Astrophysics}
\newcommand{\mnras}{Monthly Notices of the RAS}
\newcommand{\jcap}{Journal of Cosmology and Astroparticle Physics}
\newcommand{\aj}{Astronomical Journal}

\begin{abstract}
We explore a class of primordial power spectra that can fit the observed anisotropies in the cosmic microwave background well and that predicts a value for the Hubble parameter consistent with the local measurement of $H_0 = 74$ km/s/Mpc.  This class of primordial power spectrum consists of a continuous deformation between the best-fit power law primordial power spectrum and the primordial power spectrum derived from the modified Richardson-Lucy deconvolution algorithm applied to the $C_\ell$s of best-fit power law primordial power spectrum. We find that linear interpolation half-way between the power law and modified Richardson-Lucy power spectra fits the Planck data better than the best-fit $\Lambda$CDM by $\Delta$Log$\mathcal{L} = 2.5$.  In effect, this class of deformations of the primordial power spectra offer a new dimension which is correlated with the Hubble parameter.  This correlation causes the best-fit value for $H_0$ to shift and the uncertainty to expand to $H_0 = 70.2 \pm 1.2$ km/s/Mpc. When considering the Planck dataset combined with the Cepheid $H_0$ measurement, the best-fit $H_0$ becomes $H_0 = 71.8 \pm 0.9$ km/s/Mpc. We also compute a Bayes factor of $\log K = 5.7$ in favor of the deformation model. 
\end{abstract}

\pacs{PACS}
\maketitle


\section{Introduction}\label{sec:Intro}
Much work has been done in recent years to test the assumptions of the $\Lambda$CDM model ($\Lambda$ for a cosmological constant, CDM for cold dark matter).  This model, with just six parameters, has successfully explained the anisotropies in the cosmic microwave background (CMB) to a remarkable degree~\cite{Planck18cosmo}. 
Two of those parameters, $A_s$ and $n_s$, characterize the amplitude and spectral index of the primordial power spectra (PPS) of the initial Gaussian fluctuations in the early Universe.  
This parametrization is an explicit assumption about unknown physics.
There is, of course, no reason this parametrization must be true. The simplest models of inflation generically predict a power law PPS~\cite{Maldacena03,Planck13Inf,Planck15Inflation,Planck18inflation} but more complicated models can have a PPS with large deviations away from a power law, from broad features, to local ones, to oscillations~\cite{Wiggly}.  In order to test whether a power law is a sufficient parametrization for existing CMB data, or whether some alternative is needed, one can iterate over a possibly infinite number of models and check if their evidences show a strong preference for them over $\Lambda$CDM, or one can use model independent methods.  

Motivation to look for some extension to the base $\Lambda$CDM model is found in the ``$H_0$ tension''.   This tension is a disagreement between the prediction of the Hubble constant $H_0$ from the $\Lambda$CDM fit to the Planck CMB data~\cite{Planck18cosmo}, and the local measurement of $H_0$ via Cepheid calibration of supernova~\cite{Riess19}.  This tension has reached the level of $4.4\ \sigma$ and, should no potential systematic uncertainty be shown to bias either of the datasets, could point towards new physics beyond $\Lambda$CDM~\cite{Keeley19a}. A number of papers have offered a plethora of new physics explanations for this parameter discordance, though they typically involve modifying the expansion history before the surface of last scattering (early dark energy~\cite{PoulinEDE}, dark radiation, interacting neutrinos~\cite{Kreisch}), or modifying the expansion history at low redshift (evolving dark energy~\cite{Keeley19a}, dark matter interactions with dark energy~\cite{DiValentino}).  

Modifying the PPS is a potential avenue to resolve this tension that has received less attention. Previously, Hazra~et~al.~(2019)~\cite{DhirajArman} have done just this.  In their paper, they investigated what sort of PPS would be needed to explain the ``$H_0$ tension''. Hazra~et~al.~(2019) investigated possible novel PPS explanations for the $H_0$ tension by fixing the expansion history to be consistent with low-redshift observables such as the Cepheid measurement of $H_0$~\citep{Riess19} and the KiDS-450 weak-lensing measurement of $\Omega_{\rm m}$~\cite{KiDS450}.  With the expansion history fixed, they used a modified Richardson-Lucy (MRL) deconvolution algorithm \cite{Richardson,Lucy,Hazra13a,Hazra13b,Hazra14,Shafieloo04,Shafieloo07,Shafieloo08} to find what PPS maps between this fixed expansion history and the $C_\ell$s of the best-fit $\Lambda$CDM model.  They found a suppression of power at large scales and oscillations at small scales achieves this mapping and offers a potential explanation for the $H_0$ tension.

It is a somewhat generic prediction of slow-roll single-field inflation that the PPS is a nearly scale-invariant power law~\cite{BayesInflationI,BayesInflationII,Knot,Knot2}.  However, more complicated models of inflation could predict more complicated forms of the PPS~\cite{Joy,Ashoorioon,Ashoorioon2,Flauger,Achucarro}. Indeed, a power law PPS is by no means a necessary prediction of inflation. One common way to generalize the power law PPS is to allow the spectral index to vary with wavenumber, the so-called running of the index.  Similar parametrizations for the PPS such as broken power laws and steps have been explored. Similarly, steps in the inflationary potential generically give rise to non-trivial oscillating PPS.  A number of papers have explored novel inflationary potentials that give rise to oscillations in the PPS~\cite{Wiggly,2013PhRvD..87h3526A}. Thus it is natural to explore PPS beyond the power law parametrization.

In this paper, we generalize the result of Hazra~et~al.~\cite{DhirajArman} and parametrize a class of PPS that continuously deforms between the best-fit power law and the MRL-reconstructed PPS. In Sec.~\ref{sec:deform}, we elaborate on this parametrization, explaining why it fits the CMB data well, and then show the results of the statistical inference using this class of PPS in Sec.~\ref{sec:result}.  In Sec.~\ref{sec:smooth}, we describe additional ways to smooth the MRL-reconstructed PPS and in Sec.~\ref{sec:discussion} we conclude.

\section{Deformation Model}\label{sec:deform}

The primary objective of this paper is to generalize the MRL-reconstructed PPS from Hazra~et~al.~\cite{DhirajArman}, and demonstrate a class of PPS that can fit the CMB well, yet have very different expansion histories.

This sort of degeneracy between the uncertainty in the PPS and the parameters of the transfer function were explored in Kinney~(2001)~\cite{Kinney}, where it was shown that arbitrary deformations in the PPS could mimic the effects of changing the paramters of the background evolution, such as $H_0, \ \omega_{\rm b}$ or $\omega_{\rm c}$.

The idea explored by Hazra~et~al.~\cite{DhirajArman} is that, since the $C_\ell$s are the quantity directly constrained by observations, one can construct an example of a non-power-law PPS that fits the CMB exactly as well as the best-fit power law+$\Lambda$CDM model, yet has a significantly different expansion history.  In fact, Hazra~et~al.~\cite{DhirajArman} fixed the Universe's expansion history to the best-fit parameters from the Cepheid measurement of $H_0$ from Riess~et~al~(2018)~\cite{Riess18}. Further, they use the $\Lambda$CDM best-fit values for $\Omega_{\rm b}h^2$ and $\Omega_{\rm CDM}h^2$, which when combined with the Cepheid measurement of $H_0$, gives a value for $\Omega_{\rm m}$ consistent with the KiDS-450 measurement from Hildebrandt~et~al~(2016)~\cite{KiDS450}. 

With the $C_\ell$ values from the best-fit $\Lambda$CDM parameters, and a transfer function consistent with low-redshift observables, the MRL algorithm was then employed to deconvolve the transfer function from the $C_\ell$s and produce a novel PPS that predicts a high value for the Hubble parameter while still fitting the CMB well, hence offering another solution to the Hubble tension.  

To understand the effect of this PPS and why it yields a high value for the Hubble parameter, we must first understand what effect changing the Hubble parameter would have when fixing the PPS to the best-fit power law.  Just shifting $H_0$ induces a purely geometric modification to the CMB's $C_\ell$s by changing the inferred angular diameter distance to the surface of last scattering.  This, in turn, generates a phase shift in the $C_\ell$s that is consistent across the acoustic peaks.  So, in effect, the MRL deconvolution is generating a PPS that de-phase-shifts the $C_\ell$s.  

\begin{figure}
    \centering
    \includegraphics[width=\columnwidth]{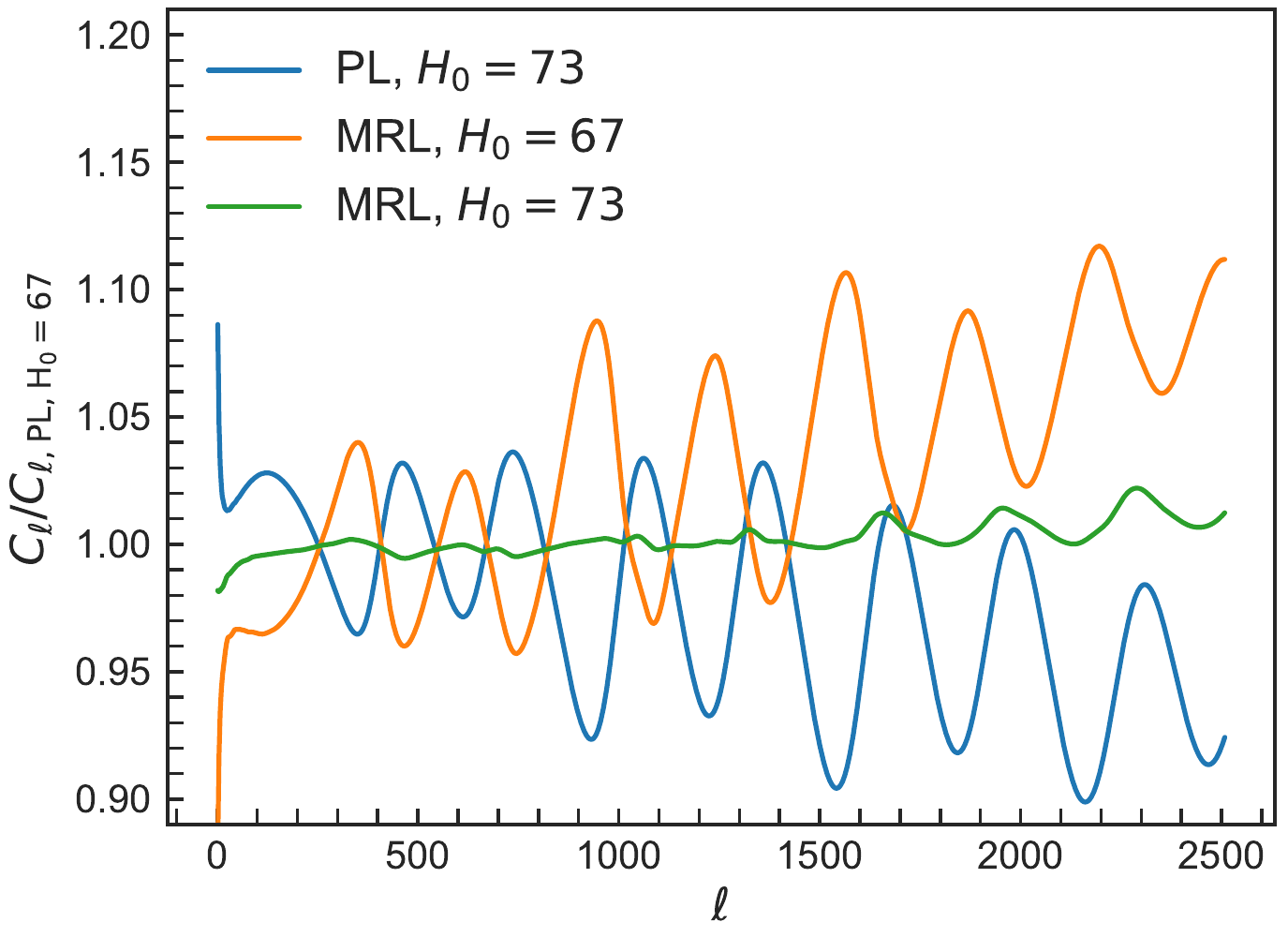}
    \caption{Ratios of $C_\ell$s relative to those from the best-fit power law PPS and background parameters.  In blue is shown the ratio of the $C_\ell$s for the same PPS but instead $H_0=73.5$.  In orange is the MRL PPS with $H_0=67.8$ and in green is the MRL PPS with $H_0=73.5$. Changing the background expansion history to $H_0=73.5$ from the one best-fit using a power law PPS effectively induces a phase shift in the acoustic peaks.  Including the MRL PPS induces the opposite phase shift, leaving the $C_\ell$s mostly unchanged. }
    \label{fig:phase}
\end{figure}

The effects, relative to the best-fit $\Lambda$CDM, of changing $H_0$, changing the PPS, and changing both, are shown in Fig.~\ref{fig:phase}.  This figure shows plots the $C_\ell$s of for a power law $\Lambda$CDM model with $H_0=73.5$ km/s/Mpc, a MRL model with $H_0=67.8$ km/s/Mpc, and a MRL model with $H_0=73.5$ km/s/Mpc, all relative to the best-fit power law $\Lambda$CDM model with $H_0=67.8$ km/s/Mpc.  The two different modifications induce equal and opposite changes to the $C_\ell$s, which when combined, induce basically no change.  

\begin{figure}
    \centering
    \includegraphics[width=\columnwidth]{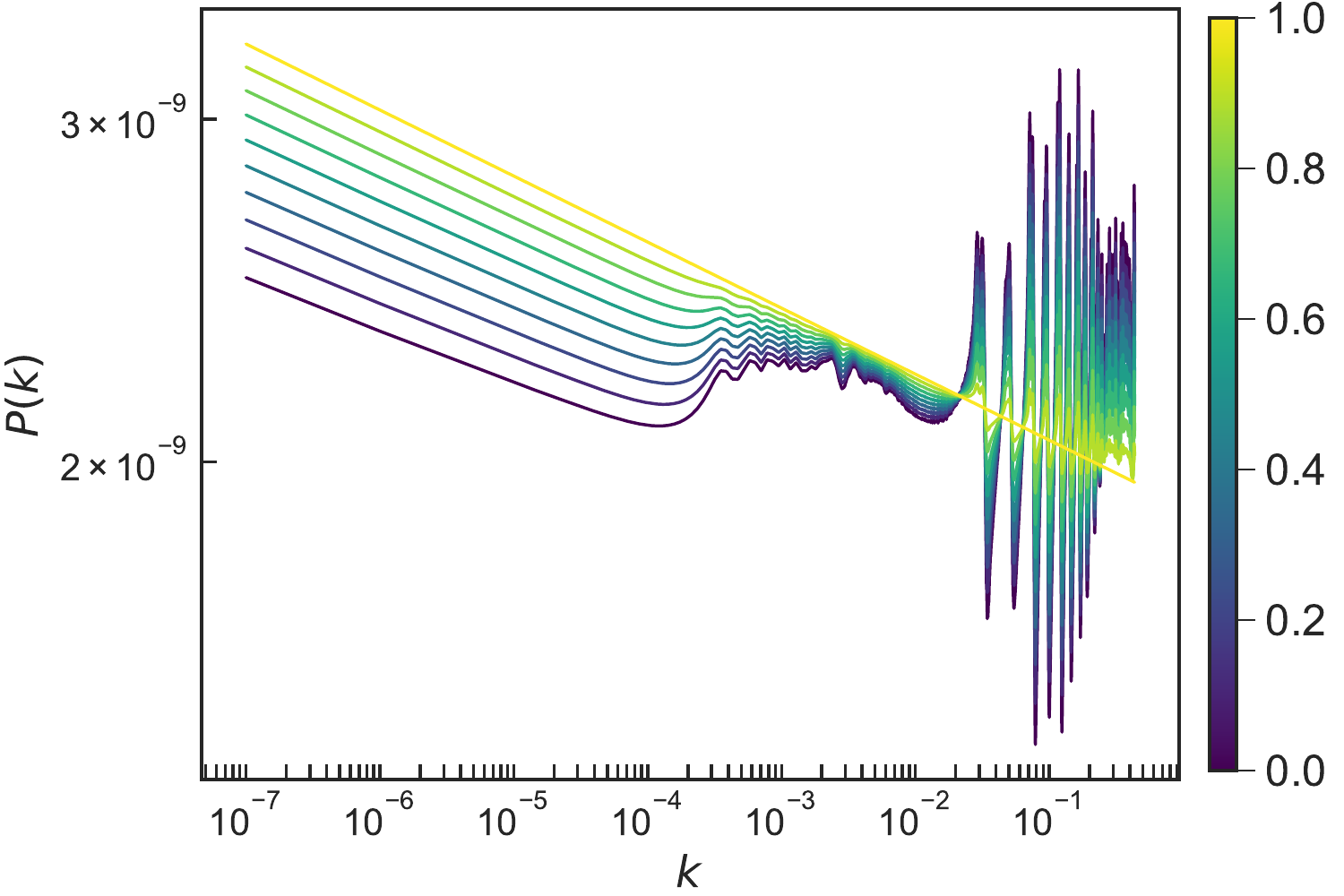}
    \includegraphics[width=\columnwidth]{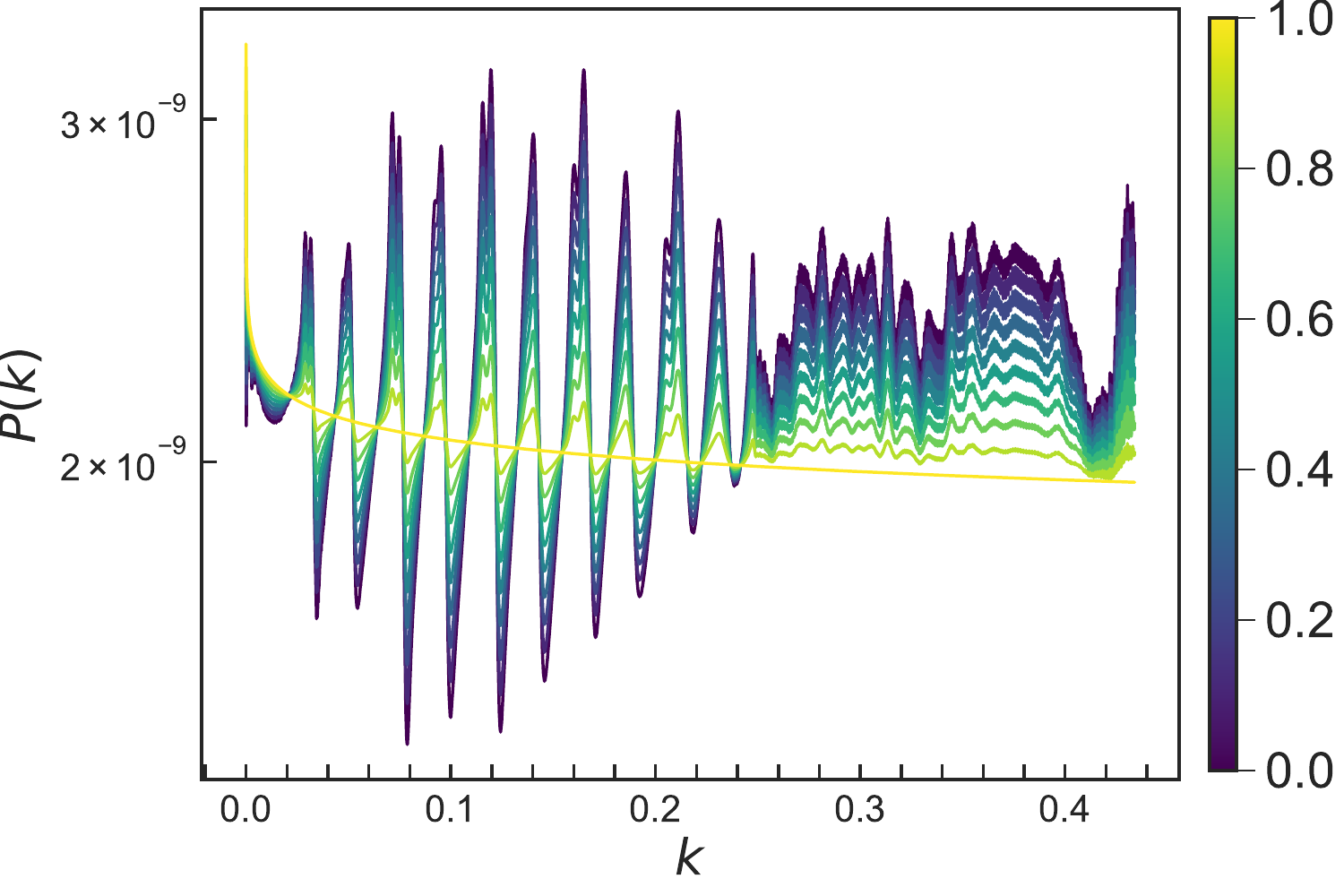}
    \caption{Example primordial power spectra that deform between the best-fit power law PPS and the MRL PPS. The viridis color map varies between purple at the MRL PPS ($f=0$) and yellow at the best-fit power law PPS ($f=1$). The top panel shows the wavenumber in log-scale and the bottom panel in linear scale, to emphasize different features of the MRL PPS.}
    \label{fig:deformations}
\end{figure}

Part of the discussion around this model must include the fact the MRL-reconstructed PPS is non-trivial and hence either potentially over-fit or a priori unlikely. Saying that this PPS is over-fit is similar to saying the MRL deconvolution used to describe was just fitting noise in the Planck 2015 data.  However, this PPS survived new additions to the dataset and changes to the modelling of foregrounds and systematics to also explain the 2018 data. That this PPS has a well-defined observable effect on the $C_\ell$s further contradicts the idea that the result is just noise.

Expressed, another way, one might reasonably believe the numerous, non-trivial features in this PPS are a priori unlikely.  Such concerns are understandable but such subjective prior belief is nothing to build firm conclusions on.  Such a prior preference for a featureless PPS lasts until someone writes down an inflationary potential that predicts the features derived in the deconvolution.  We do not seek to rule out ideas solely on a priori arguments.

In any case, the purpose of this paper is to put this MRL-reconstructed PPS on more firm ground, at least phenomenologically, by introducing a parametrized class of PPS that are a continuous deformation between the best-fit power law and the MRL-reconstructed PPS.  

This parametrization, which we refer to as the ``deformation model'' from here on, is simply an interpolation between the best-fit power law PPS and the MRL-reconstructed PPS,
\begin{equation}
    P(k,f) = P_{\rm MRL}(k) + f(P_{\rm PL}(k) - P_{\rm MRL}(k)).
\end{equation}
Thus, when $f=0$ the PPS is the MRL-reconstructed PPS, and when $f=1$, the PPS is a power law.  Fig.~\ref{fig:deformations} shows example PPS that span the space of these deformations.

In summary, we seek to perform a Bayesian model selection between the  $\Lambda$CDM model (with the base six parameters, ${\theta_s, \omega_b, \omega_c, A_s, n_s, \tau}$) and the deformation model (with those same base six parameters, along with $f$).

\section{Results}\label{sec:result}
\begin{figure*}
    \centering
    \includegraphics[width=\textwidth]{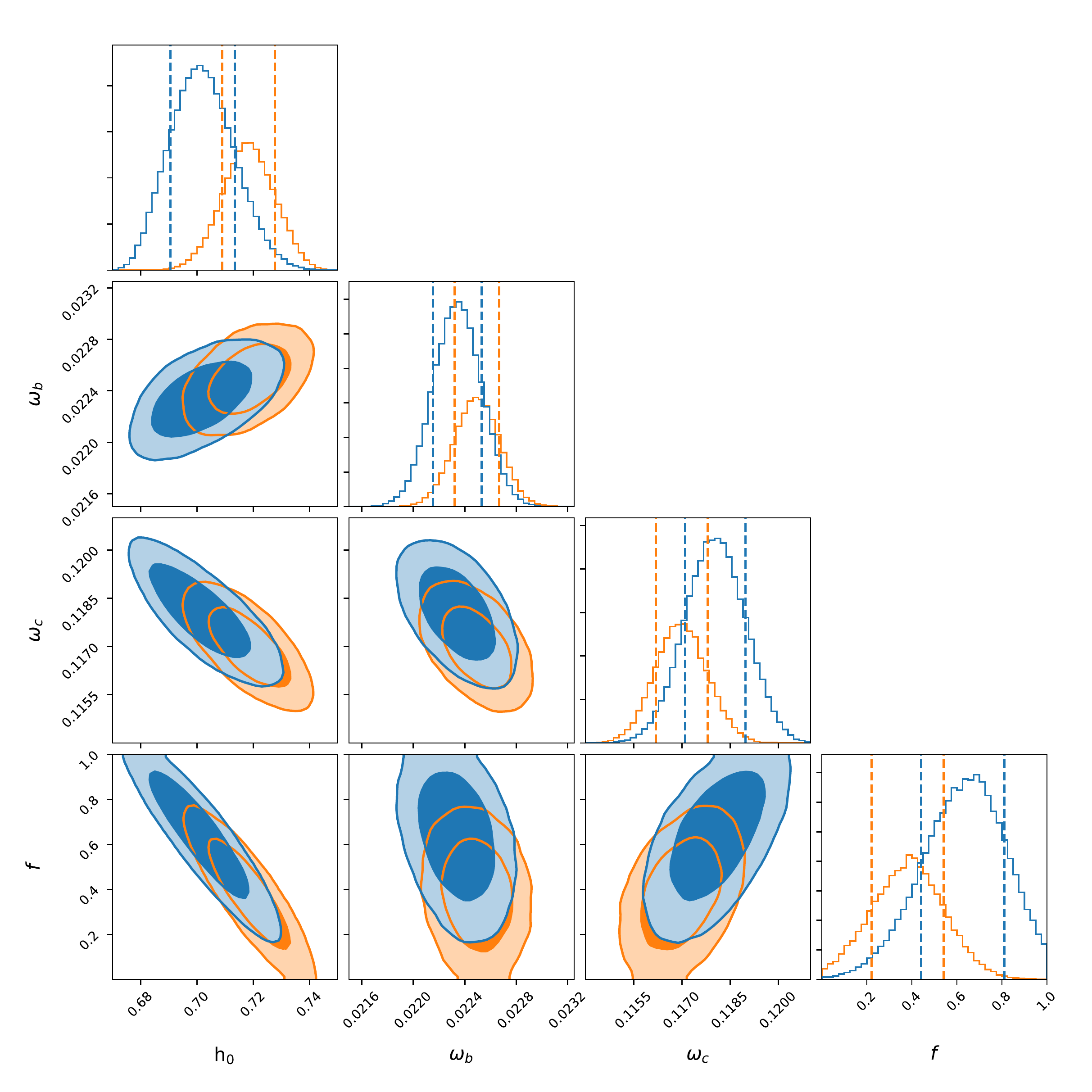}
    \caption{Posteriors for the deformation model from the Planck-TT CMB dataset (blue) and from the Planck-TT+$H_0$ datasets (orange).}
    \label{fig:corner}
\end{figure*}

First, we calculate the posteriors for the parameters of the deformation model using only the ``TT'' dataset from Planck.  This is to show, that independent of the Cepheid $H_0$ measurement, the deformation model can predict $H_0$ values higher than $\Lambda$CDM.  Then when Cepheid $H_0$ constraint is included, we show that the resulting parameter space is actually a good fit to both datasets, thus resolving the tension.  

In Fig.~\ref{fig:corner}, we show the results for our deformation model.  For the ``TT'' dataset alone, the best-fit parameters of the deformation model are $H_0 = 70.2 \pm 1.2$ km/s/Mpc and $f=0.64 \pm 0.19$.  These best-fit  parameters yield a likelihood better than the best-fit $\Lambda$CDM model of $\Delta \log \mathcal{L} = 2.5$.  This is intriguing since even on its own, the deformation model can explain the temperature anisotropies in the CMB better than the $\Lambda$CDM model, while also predicting higher values of $H_0$.  We calculate the Bayes factor for the two models ($K = Z_{\rm deform}/Z_{\Lambda\rm{CDM}}$, $Z$ is the evidence of that model) with just the Planck TT dataset to be $\log K = 2.2$

When we combine the Planck dataset with the Cepheid $H_0$ measurement from Reiss~et~al~(2019)~\cite{Riess19}, the best-fit parameters shift to $H_0 = 71.8 \pm 0.9$ km/s/Mpc and $f=0.39 \pm 0.16$ and the $\Lambda$CDM regime ($f=1$) is strongly ruled out. Further, we find the Bayes factor to be $\log K = 5.7$, which according to the Jeffreys scale~\cite{Jeffreys}, amounts to strong evidence.

\section{Smoothing}\label{sec:smooth}
In this section, we seek to answer the question of which features in the MRL-reconstructed PPS are primarily driving the preference for parameters values that are different from those inferred from $\Lambda$CDM.  One potential way to answer this question is to apply various smoothings, filters, or wavelet transforms to the MRL-reconstructed PPS and check if the resulting PPS can achieve the same likelihood.

Simple techniques like frequency cuts and low-pass filters suppress the features in the range $k\sim 10^{-4}$ to $10^{-2}$ leaving only the sinusoidal oscillations at high wavenumber. To test the effects of the filter, we scanned over the cutoff ``frequency'' of the low-pass filter and calculated how much the likelihood changed. For larger values of the cutoff frequency, the change in the PPS is smaller.  When scanning over the cuttoff frequency, we found no preference for any filtering but some amount was still allowed. Applying a cutoff frequency of 0.6 Mpc to the MRL PPS, the likelihood is marginally worse ($\Delta \log \mathcal{L} \sim 0.5$), but the jaggedness in the high-wavenumber part of the MRL PPS is removed.  The cost in the likelihood  buys a priori subjective belief. Further, all of the low-wavenumber features are also removed.  

\begin{figure}
    \centering
    \includegraphics[width=\columnwidth]{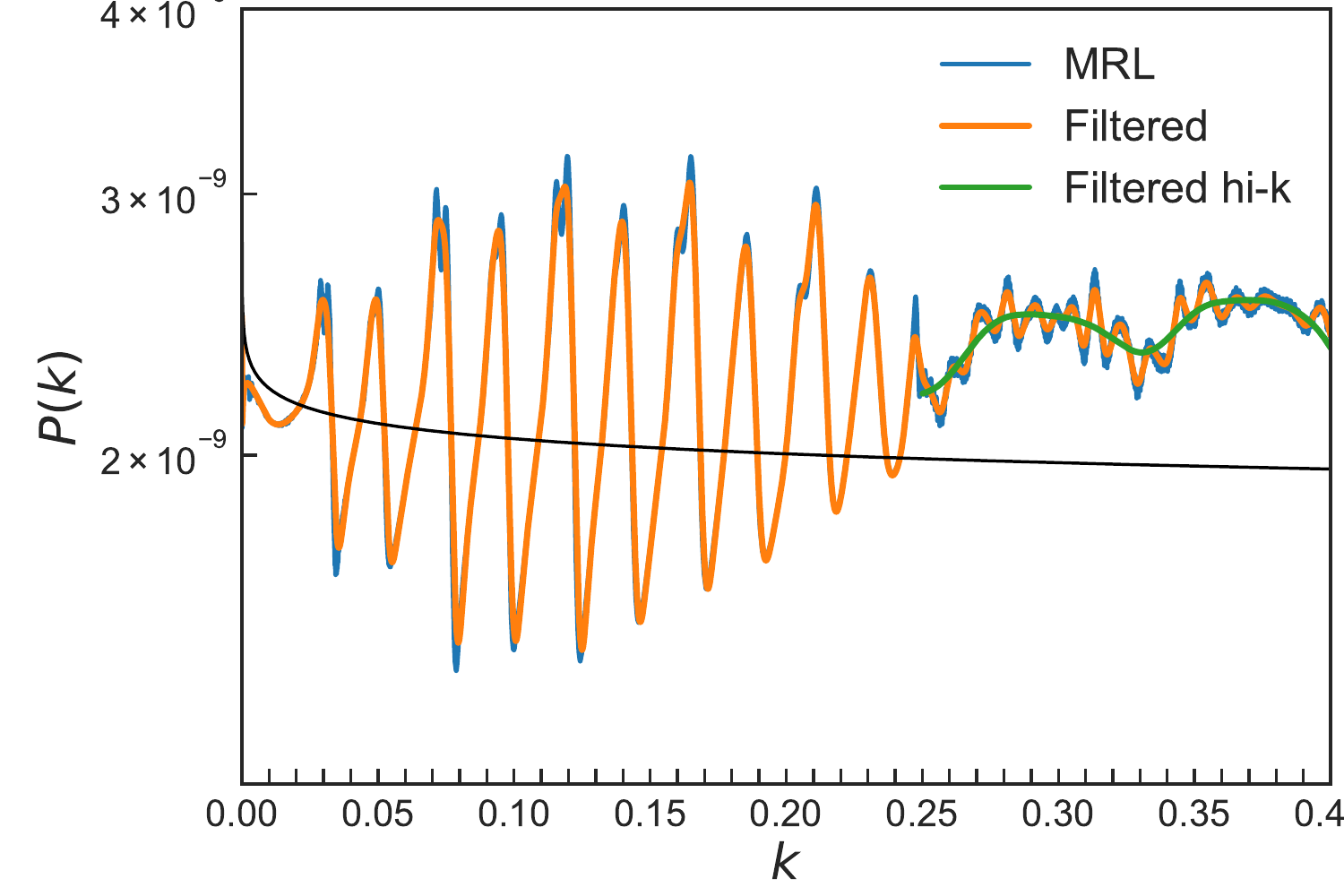}
    \caption{Example filtered MRL PPS, where the highest frequency variations in the MRL PPS were suppressed (blue).  The raw MRL PPS is shown in orange.}
    \label{fig:smooth}
\end{figure}

To more precisely answer the question of which features give rise to the MRL preference for a high $H_0$, we also tested a case where the low-pass filter was applied to only the features above $k>0.25$. This choice was motivated by the fact that above $k>0.25$ only meaningfully affect the $C_\ell$ values for $\ell > 2500$, which is a regime that is unprobed by CMB observations. Keeping with these expectations, we find that essentially all of the features above $k>0.25$ can be filtered away and maintain the same likelihood.  We show this ``hi-k'' filtering case in Fig.~\ref{fig:smooth} where in green, we show the filtered part of the PPS (the blue MRL and the green filtered curves combine to form the tested PPS).
Thus, it is apparent that these mid-wavenumber ($0.01<k<0.25$) oscillations are what compensate for the phase shift in the acoustic peaks coming from the different background parameters.

It is not unreasonable that these features could arise from steps or kinks in a physical inflationary potential such as wiggly-whipped inflation~\cite{Joy,Wiggly,2013PhRvD..87h3526A}.  Thus these sort of ``deformed'' or ``filtered'' PPS models offer a new hope that inflationary physics beyond the simplest single-field slow-roll inflation might be true.

\section{Conclusions}\label{sec:discussion}

We generalized the results of Hazra~et~al.~(2019)~\cite{DhirajArman} and introduced a class of PPS that can explain the temperature anisotropies in the CMB well, yet also predict high values for the Hubble parameter $H_0$ (i.e. consistent with the Cepheid measurement).  This generalization is simply an interpolation between the MRL-reconstructed PPS and the best-fit power law PPS, which we call the deformation model.  

We performed a Bayesian analysis to compare the base $\Lambda$CDM model and the deformation model and then calculate the posterior for the parameters of that model. We find that the deformation model correlates $H_0$ with the new degree of freedom in the PPS, thus predicting higher values for $H_0$ from the CMB's TT dataset than the base $\Lambda$CDM model.  This is not just the uncertainties on $H_0$ increasing to alleviate the tension with the Cepheid $H_0$ measurement from Riess~et~al~(2019)~\cite{Riess19}, but the best-fit values shift towards higher values ($H_0 = 70.2$ km/s/Mpc), even without the Cepheid constraint.  When the CMB's TT dataset is considered jointly with the Cepheid $H_0$ measurement, we find that the deformation is preferred over $\Lambda$CDM by a Bayes factor of $\log K = 5.7$. 

Additionally, we have explored the question of which features in the MRL-reconstructed PPS are driving the preference for different parameter values.  Simply put, it is the features at intermediate wavenumber $k \sim 0.01 -- 0.25$ are most important for fitting the acoustic peaks and hence the parameters of the background expansion.  The features at high-$k$ ($k>0.25$) can be replaced with a power law to recover the same $C_\ell$s and likelihood of the best-fit $\Lambda$CDM parameters.  

The most important conclusion to take away from this work is that there exist unaccounted for degeneracies between the uncertainties in the PPS and the background expansion history.  Even beyond the MRL-reconstructed PPS, arbitrary deformations of the PPS can mimic changes arising from different expansion histories.  Though whether these classes of deformed PPS are a priori unreasonable is still an open question, we have shown that the deformation model is a posteriori reasonable.  Even if one were skeptical that a physically motivated model of inflation could ever generate a PPS with the features in the MRL-reconstructed PPS, this work is still useful to demonstrate how far and in what ways one would have to deform a power law PPS to beat the successes of $\Lambda$CDM. Because we are primarily interested in data driven techniques, we find it intriguing that deformations of the PPS can be correlated with the parameters of the background expansion. This makes inferences from the CMB less certain, and it is important to remain agnostic and open about these ideas especially considering the Hubble tension. 

We leave the calculation of the posteriors of the deformation model from the polarization data and additional low-redshift probes for future work.


\bibliography{sample}

\end{document}